%% file: main.tex
\documentclass[conference]{IEEEtran}
\IEEEoverridecommandlockouts
\usepackage{cite}
\usepackage{amsmath,amssymb,amsfonts}
\usepackage{colortbl}
\usepackage{algorithmic}
\usepackage{graphicx}
\usepackage{textcomp}
\usepackage{xcolor}
\usepackage{hyperref}
\usepackage{ stmaryrd }
 \usepackage[T1]{fontenc}
 \usepackage{enumitem}
 \usepackage{balance}
\usepackage{microtype}
\usepackage{comment}
\usepackage{listings}
\usepackage{placeins}

\usepackage[numbers]{natbib}
\usepackage{mathtools}

\usepackage[linesnumbered, ruled]{algorithm2e}
\SetAlCapSkip{1em}
\SetKwInput{KwParam}{Parameters}
\SetKwInput{KwHyper}{Hyperparameters}

\SetCommentSty{mycommfont}
\usepackage{cleveref}
\usepackage{bm}
\usepackage{float}
\usepackage{fancyhdr}

\newcommand{\linebreakand}{%
  \end{@IEEEauthorhalign}
  \hfill\mbox{}\par
  \mbox{}\hfill\begin{@IEEEauthorhalign}
}

\newcommand{\name}[0]{\texttt{SLaDe}} 

\def\BibTeX{{\rm B\kern-.05em{\sc i\kern-.025em b}\kern-.08em
    T\kern-.1667em\lower.7ex\hbox{E}\kern-.125emX}}

\begin{document}

\title{\name{}: A Portable Small Language Model Decompiler for Optimized Assembly}


\author{\IEEEauthorblockN{Jordi Armengol-Estapé}
\IEEEauthorblockA{\textit{School of Informatics} \\
\textit{University of Edinburgh}\\
Edinburgh, United Kingdom \\
jordi.armengol.estape@ed.ac.uk}
\and
\IEEEauthorblockN{Jackson Woodruff}
\IEEEauthorblockA{\textit{School of Informatics} \\
\textit{University of Edinburgh}\\
Edinburgh, United Kingdom \\
j.c.woodruff@sms.ed.ac.uk}
\and
\IEEEauthorblockN{Chris Cummins}
\IEEEauthorblockA{
\textit{Meta AI Research}\\
Menlo Park, CA, USA \\
cummins@fb.com}
\and
\IEEEauthorblockN{Michael F.P. O’Boyle}
\IEEEauthorblockA{\textit{School of Informatics} \\
\textit{University of Edinburgh}\\
Edinburgh, United Kingdom \\
mob@inf.ed.ac.uk}
}
\maketitle

\thispagestyle{fancy}

 
Decompilation is a well-studied area  with numerous high-quality  tools available. These are frequently used for 
security tasks and to port legacy code. However, they regularly
generate difficult-to-read programs
and require a large amount of engineering effort to support new programming languages and ISAs. 
Recent interest in neural approaches
has produced portable tools that generate readable code.
Nevertheless, to-date such
techniques are usually restricted to synthetic programs without optimization, and no models have evaluated their 
portability. Furthermore, while
the code generated may be  more readable, it is usually incorrect.

This paper presents \name{}, a  {\bf S}mall {\bf La}nguage model
{\bf De}compiler based on a sequence-to-sequence Transformer trained over real-world code and augmented with a type inference engine.
We utilize a novel tokenizer, dropout-free regularization, and type inference
to generate programs
that are more readable and accurate than standard analytic and  recent neural approaches. Unlike standard approaches, \name{} can infer out-of-context types and unlike neural approaches, it generates correct code.

We evaluate \name{} on over 4,000 ExeBench
functions on two ISAs and at
two optimization levels. \name{} is up to
$6\times$ more accurate than Ghidra, a
state-of-the-art, industrial-strength decompiler
and up to $4\times$ more accurate than
the large language model ChatGPT and generates
significantly more readable code than both.


\input{introduction}
\input{motivation}
\input{translation}
\input{pretrain}
\input{train}

\input{inference}
\input{setup}
\input{results}

\input{analysis}

\input{related}
\input{conclusion}



\section*{Data Availability Statement}

We refer to the accompanying artifact \cite{slade_authors_2023_10205121}.

\section*{Acknowledgements}

We thank Irina Rish, supported by the Canada CIFAR AI Chair Program and the Canada Excellence Research Chairs Program, and Compute Canada, for the help with part of the compute. We thank the reviewers for their insightful comments. 



\bibliography{main}
\bibliographystyle{IEEEtran}
\balance
\input{appendix}

\end{document}

%% file: introduction.tex
\section{Introduction}
\label{sec:intro}



Decompilation is the automatic lifting of
assembly instructions to  higher-level, human-readable source code~\cite{cifuentes1995decompilation}.
Decompilation enables the porting of legacy programs to new hardware \cite{stitt2008binary}, and 
the detection of malware for security protection \cite{Hosseini2022}.
50 years of decompilation research~\cite{housel1973study}
has resulted in a number of commercial~\cite{retdec}
and open-source tools~\cite{ghidra}, which represent many years of engineering
effort~\cite{DBLP:journals/corr/abs-1905-08325}.

Although well-studied, decompilers 
are fundamentally limited by their inability to determine the types of variables and functions
declared  outside the source assembly, relying on programmer assistance.  
In terms of readability, 
they 
rely on large bodies of pattern-matching
rules~\cite{hex}, 
which 
generate hard-to-read code. 
Work to make these
decompiled codes less complex~\cite{chen2013refined,ghidra, yakdan2015no}
is still well short
of what a programmer would write.

Neural decompilation~\cite{8330222} promises to overcome
this
producing
more human-readable programs~\cite{DBLP:journals/corr/abs-2107-03374},
but existing techniques usually rely on synthetic
data and do not generate correct code.
Neural decompilation techniques \cite{DBLP:journals/corr/abs-1905-08325} rely on token or string-level similarity metrics, such as edit-distance \cite{Hosseini2022}
which are effective in natural language translation \cite{maarif2014complexity}.
However,
these metrics 
fail to  capture  the 
correctness of a translation. 
%
The result is generated code that looks similar to the ground-truth but is incorrect.

Recently, large-language models such as
ChatGPT~\cite{chatgpt} have
received considerable
attention.
These models are able to generate high-quality programs
from text inputs, at the cost of extreme model size.
However, as we show 
in section \ref{sec:results}, ChatGPT performs poorly 
at decompilation, frequently producing incorrect and
hard to read code.


In summary, decompilers fall into two categories:
traditional hand-crafted decompilers that produce hard
to read code and fail to tackle external type declarations, 
and neural techniques that produce incorrect answers.
What we want is a technique with
the accuracy of traditional schemes that produces
code as readable as neural approaches,
works on real-world  code, manages external type declarations and is
portable across ISAs and code optimization levels
with minimal engineering effort.

\subsection{Summary of Approach}
We present {\tt SLaDe}, a {\bf S}mall {\bf La}nguage model
{\bf De}compiler. It comprises a 200M-parameter Transformer specifically trained for decompilation at program function, rather than code fragments, scale. We 
develop a novel code tokenizer and train a sequence-to-sequence,  assembly to
C model with no dropout. 
Because \name{} uses a neural approach, we generate readable code
compared to existing approaches (Section~\ref{sec:EditDistance}) and have easy portability across
ISAs and optimization levels.  While
in-principle other neural approaches
are portable, all existing work targets 
x86 exclusively.
\name{} is the first neural decompiler to be applied across ISAs and optimization levels.

To decompile code, \name{} applies
the trained model to assembly.  This
produces C code, which often contains
undefined types, similar to existing
rule-based decompilers.
We use PsycheC~\cite{Melo2018} to
generate types for the resulting source
code (Section~\ref{sec:TypeInferenceAlgorithm}).
This combination of neural translation and compiler analysis is crucial to decompilation accuracy (Section~\ref{sec:analysis}).

We perform a large-scale evaluation on over 4,000
executable programs from ExeBench \cite{exebench} (following AnghaBench \cite{angha} methodology for obtaining compilable C functions, and ExeBench methodology for making them executable) and compare against the neural decompiler BTC  \cite{Hosseini2022}, the state-of-the-art industrial strength decompiler Ghidra \cite{ghidra} from the NSA, and the large language model ChatGPT~\cite{chatgpt}. We show that it generates code that is both more accurate and more readable than all existing schemes across ISA and optimization level.
We are up to 6$\times$ more accurate than Ghidra and  up to 4.2$\times$  more accurate
than ChatGPT. In addition, we generate code that is 1.65$\times$ to 3.68$\times$ more edit-similar to the original source  than Ghidra and 1.07$\times$ to  3.83$\times$ nearer
than ChatGPT.

This paper makes the following contributions:
\begin{enumerate} 
    \item We present the first  neural decompiler that works across ISAs and optimization levels.
    \item We introduce type inference-augmented neural decompilation, together with 
    novel tokenization, dropout-free
    training over real-world code which
    dramatically improves semantic accuracy compared to existing neural decompilers.
    \item \name{} outperforms  industrial strength rule-based approaches in terms of accuracy  while significantly improving readability.
    \item We show in a large-scale evaluation on a dataset of 4,000 functions across two optimization levels and two ISAs 
    that our small  model significantly outperforms ChatGPT with three orders of magnitude fewer weights. 
\end{enumerate}     

%% file: motivation.tex
\section{Motivation}
\label{sec:motivation}
This section presents an example illustrating the 
complexity of decompilation
before sketching an outline of our approach.

\begin{figure*}[thbp!]
\includegraphics[width=\textwidth]{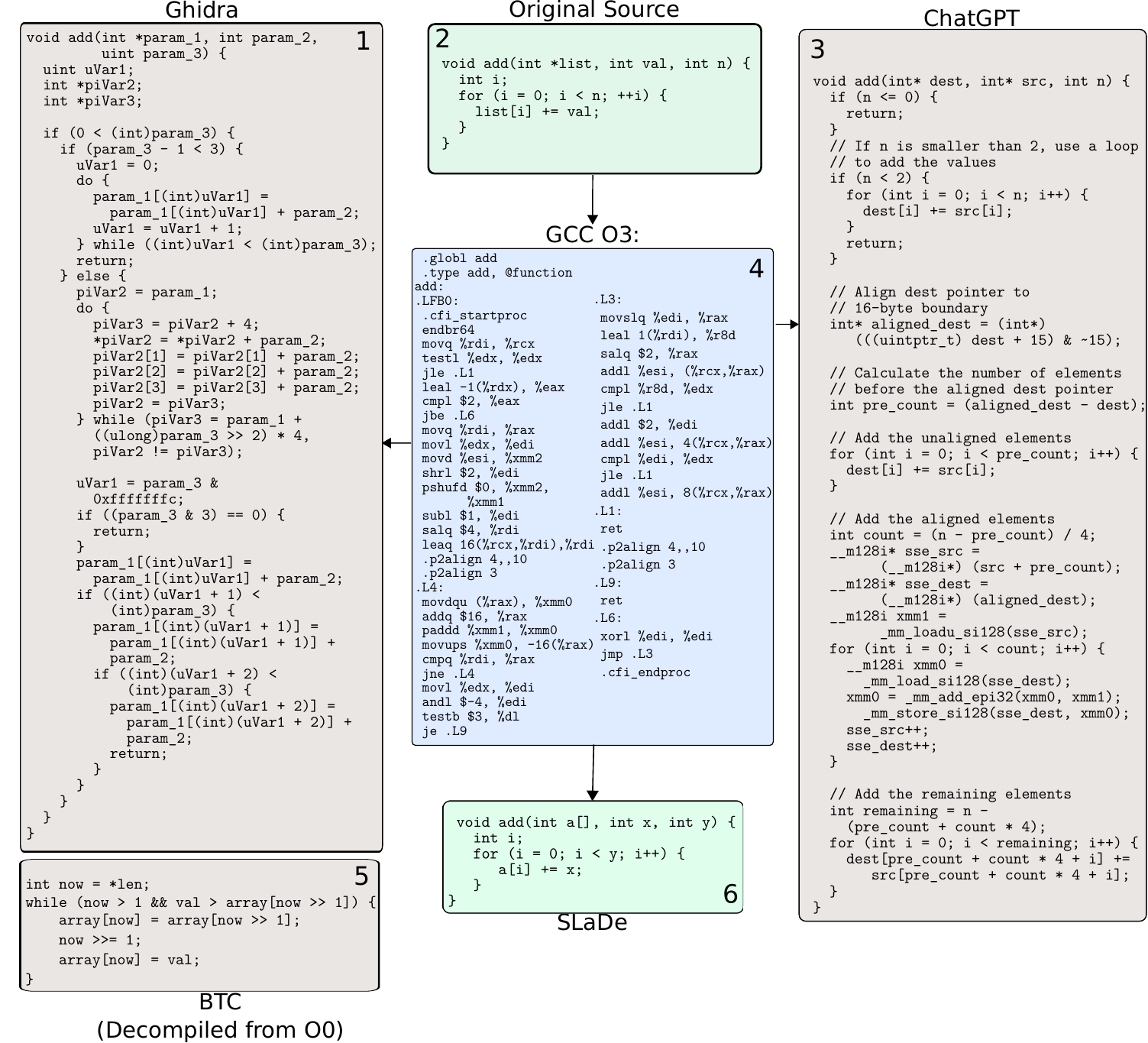}
    \caption{Comparing
    decompilation techniques to the ground-truth.  We compile the original code (box 2) using GCC O3 and then decompile with each technique.  As BTC was trained on O0 code, we use O0 to evaluate it.  We
    can see that Ghidra (box 1) and ChatGPT (box 3) produce very difficult to read
    code and in ChatGPT's case, the code is incorrect, adding two arrays
    together rather than
    adding a constant to an array.
    BTC (box 5) produces significantly more readable, but also incorrect, code.  \name{} (box 6) produces readable, correct code. }
    \label{fig:DecompileExample}
\end{figure*}
\subsection{Examples}

Figure~\ref{fig:DecompileExample} shows a typical decompilation task selected from the Synth Benchmarks 
 where Box 2 contains  the original
C code. 
If we compile this function with 
GCC using  level -O3 optimization, we
obtain the assembly in Box 4.
GCC's -O3 optimization has added significant complexity, unrolling and vectorizing the loop resulting in 55 lines of assembly, multiple loops
and vector instructions.
By feeding this assembly code into various decompilers we receive the following outputs:

\emph{Ghidra}~\cite{ghidra}, a state-of-the-art, industrial strength, patter-matching  decompiler generates the code shown in Box 2; a literal translation of the input assembly into a higher-level source language. While correct, the generated code is tightly coupled to the input assembly's structure, making it verbose, hard to read, and bears little resemblance to the input source code. Ghidra makes no attempt to produce interpretable variable names.

\emph{BTC}~\cite{Hosseini2022}, the only publicly-available neural decompiler that works on non-synthetic code, aims to reduce the readability gap of traditional decompilers. BTC accepts only unoptimized assembly so we fed as input the -O0 assembly (not shown). While the output, shown in Box 5, is less verbose and more readable than Ghidra, it does not contain a function header and has incorrect behavior.

\emph{ChatGPT}~\cite{chatgpt} improves over BTC in that it generates a complete function definition as shown in Box 3. It infers human-readable variable names and even adds comments.
However, like Ghidra, the output is long and has complex control-flow. Surprisingly, 
it contains x86 intrinsic instructions 
obscuring the meaning of the function. 
Furthermore, when executed it gives incorrect results.

\emph{\name{}}, presented in this paper, produces output with correct control-flow structure and the same number of variables as shown in Box 6. While the names of the variables are different and the pointer argument is equivalently declared as an array, the behavior of this code is immediately obvious and is functionally equivalent to the original.

\subsection{Our Approach}
\begin{figure*}[t]
\includegraphics[width=\textwidth]{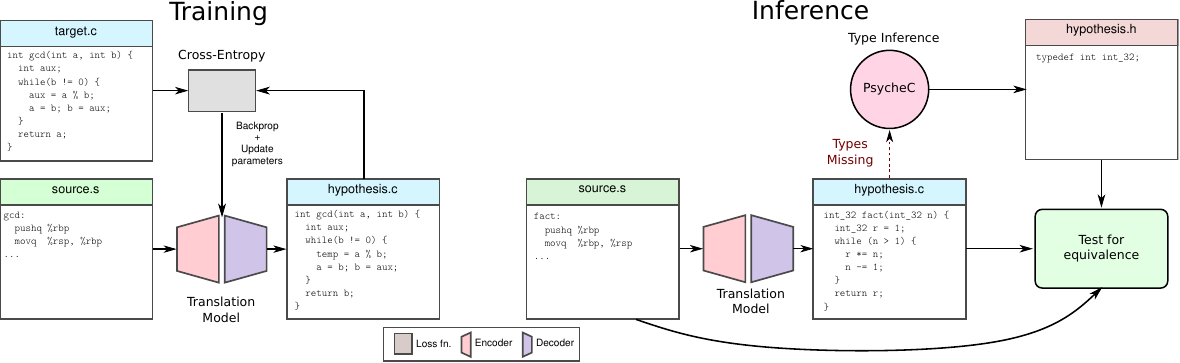}
\caption{
We train a small Transformer to minimize the cross-entropy loss
function.
At inference time, we use
the model to generate
code.  Generated code with missing typedefs
is passed to PsycheC~\cite{Melo2018} to generate
candidate types.  We then check the inputs
for correctness using input/output examples.
}
\label{fig:DecompilerDiagram}
\end{figure*}

Figure~\ref{fig:DecompilerDiagram} gives an overview of our approach.
We represent programs as a sequence of tokens and use a sequence-to-sequence encoder-decoder Transformer model.
We train the model with (assembly, C) sequence pairs, adjusting weights to minimize the difference
between the predicted output and the correct one
using the cross-entropy loss function.  Unlike standard approaches,
we do not use dropout regularization \cite{JMLR:v15:srivastava14a}. As per our preliminary experimentation, weight decay regularization alone (with no dropout) yielded better results.

During inference, we run the model over
the source assembly, using beam search to generate
a C program.  We apply type inference (Section~\ref{sec:TypeInferenceAlgorithm}) to
generate missing types required to run generated
program if it is required.  
Provided the decompiled program is compilable and
executable,
we test it for equivalence using I/O examples (Section~\ref{sec:IOCorrectness}).

In this work, we tackle the task of decompilation from assembly instructions to C, not disassembly. That is, we directly feed assembly instructions to \name{} rather than machine bytes. The task of high-quality disassemblers such as \cite{dis}, while complementary to our work, is out of the scope of this article.

%% file: translation.tex
\section{Decompilation as neural machine translation}
\label{sec:method}

We learn a model $T$  that takes as input a low-level assembly program as input and outputs 
a high-level C program.
Our model supports optimized
an unoptimized assembly, and
handles x86 and Arm assembly.

\begin{equation}T: S \mapsto C \end{equation}

where $S = \text{x86} \cup  \text{Arm}$, the set of all x86 and Arm assembly
programs at optimization levels O0 and O3, and $C$ is the set of
all C-language programs.

We represent each instance of ${S, C}$ as
a sequence of tokens.
$c= [c_1, \ldots c_{n} ] \in C  ,  s= [s_1, \ldots s_{m}] \in S$
and restrict program length $m=1024$.

\subsection{Equivalence}
\label{sec:correct}
For correct  translation between some assembly
$s \in S$ and some source $c \in C$,
we require that  meaning of $s$ and $c$ are identical
\begin{equation}M\llbracket{s}\rrbracket \equiv M' \llbracket{c}\rrbracket 
\label{EquivalenceEquation}
\end{equation}

Where $M$, $M'$ are functions denoting the meaning
of programs in $S$ and $C$ respectively.  We use
$\equiv$ to denote equivalence of program state rather
than just return values.
In general equation~\ref{EquivalenceEquation} is undecideable~\cite{Winskel1993} and  we instead consider  both more restricted and looser
bounds of equivalence, Input/Output Equivalence
on finite subsets.

\paragraph{Input Output (IO) Accuracy}
\label{sec:IOCorrectness}
A typical definition of semantic equivalence is to use
behavioural equality.
That is, for some functions $c \in C$ and $s \in S$
over some domain $D$:
$\forall x \in D. s(x) \equiv c(x)$
. This is undecidable for domains where $D$ is infinite, which is frequent in practice.  To make
this notion of equivalence decidable, we instead use input/output
equivalence on \textit{finite subsets}.  That is,
we select some finite set $\mathcal{F} \subseteq D$ and then seek
to prove that:
\begin{align}
    \forall x \in \mathcal{F}. s(x) \equiv c(x)
\end{align}

This process is decidable provided we assume that $c$ or $l$
terminates.  In practice, non-termination is rare and we
assume non-equivalence in cases on non-termination.
This process used for compiler testing in~\cite{le2014compiler} is successfully used in evaluating standard decompilers in~\cite{liu2020far}; this paper is the first to use it on neural decompilation. 
As the size of the finite subset increases, our confidence increases that the programs  are truly equivalent also increases, but is only guaranteed in cases where $D$ is finite and fully-explored.
In many case tighter bounds on equivalence can be found via model-checking and other techniques~\cite{dasgupta2020scalable}.

\paragraph{Edit similarity}
\label{sec:EditDistance}
A key metric for decompilation tools is the readability
of the code.  Readbility is of course a subjective metric,
but for a decompilation task, we are most interested  in similarity  to  the original source code.
We use edit distance, a standard metric in   other neural  approaches~\cite{katz2019neural,Hosseini2022} to give  \textit{edit similarity} a measure of 
closeness. 

The edit \textit{distance} is defined as the minimum
number of \textit{insertion},
\textit{deletion}, and \textit{replacement} operations
needed to transform one string into another.
Given some decompiled C, $c^* = \{c^*_1, \dots, c^*_n\}$
and a ground-truth source C, $c = \{c_1, \dots, c_n\}$,
We can compute the edit distance using a dynamic-programming
based algorithm defined in figure~\ref{EditDistanceFigure}.  We use
$\varepsilon$ to represent
the empty sequence.
We use \textit{edit similarity},
is $1 - \frac{\text{Edit Distance}}{  \text{Sequence Length}}$, which
is normalized to sequence length (of the ground truth target)
and converted so that
a higher edit similarity represents
better readability.


\begin{figure}
\begin{flalign*}
    \text{Distance}&(\varepsilon, \{c_1, \dots, c_j\}) = j \\
    \text{Distance}&(\{c^*_1, \dots, c^*_k\}, \varepsilon) = k \\
    \text{Distance}&(\{c, c^*_2, \dots, c^*_k\}, \{c, c_2, \dots, c_j\}) = \\
    &\text{First tokens equal:}\\
    &\text{Distance}(\{c^*_2, \dots, c^*_k\}, \{c_2, \dots, c_j\}) \\
    \text{Distance}&(\{c^*_1, c^*_2, \dots, c^*_k\}, \{c_1, c_2, \dots, c_j\}) = \\
    &\text{One case for each of delete, insert and replace:} \\
    &\min(\\
    &\quad\text{Distance}(\{c^*_2, \dots, c^*_k\}, \{c_1, c_2, \dots, c_j\}) + 1, \\
    &\quad\text{Distance}(\{c^*_1, c^*_2, \dots, c^*_k\}, \{c_2, \dots, c_j\}) + 1, \\
    &\quad\text{Distance}(\{c^*_2, \dots, c^*_k\}, \{c_2, \dots, c_j\}) + 1 \\
    &)
\end{flalign*}
\caption{
Algorithm for computing the edit-distance between
two sequences.  We use
$\varepsilon$ to represent
the empty sequence.
We use \textit{edit similarity},
which is $1 - \text{Edit Distance} / \text{Sequence Length}$, so that
a higher edit similarity represents
better readability.
}
\label{EditDistanceFigure}
\end{figure}

%% file: pretrain.tex

\section{Tokenization}  

Our model consumes and produces sequences of tokens from a fixed vocabulary. Since we also model identifiers, this could cause encountering unknown tokens during inference. To avoid those out-of-vocabulary tokens, we use subword tokenization~\cite{sennrich-etal-2016-neural}, in which tokens are derived from the character frequencies in the train set. Because individual characters present in the train set (in this case, essentially the ASCII alphabet) are also part of the vocabulary, unseen tokens can always be built from seen subwords, even character by character if required. We base our subword tokenization on UnigramLM~\cite{https://doi.org/10.48550/arxiv.1808.06226}, which has been shown to either match or exceed (by up to 10 F1-score points in some downstream tasks~\cite{bostrom-durrett-2020-byte}) the performance of Byte-Pair Encoding \cite{sennrich-etal-2016-neural}.

We modify the default parameters of UnigramLM to make it more amenable to code. We set a small vocabulary size of 8k due to the small number of C keywords and assembly opcodes as compared to unrestricted natural language (vocabulary sizes in natural language processing are typically >30k). We tokenize numbers digit-by-digit as in~\cite{muffo-etal-2022-evaluating}, e.g. $512 \rightarrow [5,1,2]$. This prevents inconsistencies when encoding large numbers (e.g.\ $512 \rightarrow \{[5,1,2], [5,12],[51,2],[512]\}$. We split all punctuation signs into different tokens (so that dots are not merged in float numbers, etc). We protect spaces by escaping with the metaspace character (unicode \_) as in SentencePiece~\cite{sentencepiece_}, but only inside double quotes for string definitions. Otherwise, spaces are normalized (replaced with a single space).



%% file: train.tex
\section{Training}

We pose neural decompilation as a sequence-to-sequence task. Given a dataset $\mathcal{D}$ with $N$ pairs $\{(\mathbf{s_n}, \mathbf{c_n}) | n \in \{0..N-1\}\}$, where $\mathbf{s_n}$ is a function assembly code and $\mathbf{c_n}$ is the corresponding C code that produced it, we want to train a model T parameterized by $\bm{\theta}$ with maximum likelihood estimation:

\begin{align}
\bm{\theta}^{*} = \arg\max_{\bm{\theta}} \prod_{n=0}^{N-1} P(\mathbf{c_n} | \mathbf{s_n}; \bm{\theta})
\end{align}

In practice, we minimize the negative log-likelihood, which is equivalent to (multiclass) cross-entropy (CE). For each minibatch (i.e., group of $B << N$ examples in the dataset), we use stochastic gradient descent to update the parameters of the model given the gradient of the cross-entropy loss function:


\begin{align} 
\boldsymbol{P}(\boldsymbol{\theta}) = \operatorname{T}\left(\boldsymbol{s}_n, \boldsymbol{c}_n \mid \boldsymbol{\theta}\right)
\end{align}

\begin{align}  \operatorname{CE}(\boldsymbol{\theta})=-\sum_{l=1}^{L-1} \log P(\boldsymbol{\theta})\left[\boldsymbol{c_n}[l+1], l\right] 
\end{align}
\begin{align} \boldsymbol{\theta'} = \boldsymbol{\theta}-\eta \cdot \nabla \operatorname{CE}(\boldsymbol{\theta})
\end{align}

where $\boldsymbol{P}(\boldsymbol{\theta})$ are the token probabilities predicted by the model (the expected outputs $\boldsymbol{c}_n$ are also needed by the model in training due to the use of teacher forcing\footnote{That is, in training, unlike in inference, when predicting the token $l$, the model is fed the ground truth up to $l-1$, not its own predictions autoregressively. This means that in training the decoder can run in parallel, and that compute is not thrown away by trying to predict sequences that are already too far off from the ground truth.}),  $L$ is the sequence length, $\eta$ is the learning rate hyperparameter, and $\boldsymbol{\theta'}$ are the updated parameters.




\subsection{Training dataset}

AnghaBench \cite{angha} is dataset of compilable C functions scraped from public repositories. We use an expanded version with around 4M functions paired with the corresponding function-level assembly obtained from ExeBench \cite{exebench}. The scraped functions contain no restrictions so that the code is representative of real-world code.
We ensure that the test set functions
are not present in the training set with token-level hash-based deduplication.
We feed functions (as opposed to programs) to our model to avoid long sequences, which
would make the learning task more challenging and more
computationally expensive.

The assembly-C pairs come from GCC with different optimization
levels.
For each assembly-C pair, we take the assembly function without
its surrounding context and ask our model to predict the corresponding C function without any surrounding context.
By design, \name{} does not have access to functions, typedefs
or variables external to the function.  These are predicted
using type-inference (section~\ref{sec:TypeInferenceAlgorithm}).
For our evaluation, we
add the original context of the C program into the context
we evaluate it in.  We use these inferred types for compilation and execution
of hypothesis decompilations.  

\subsection{Sequence-to-Sequence Transformer}


We use a sequence-to-sequence Transformer model \cite{Vaswani2017}. Each input token from the source sequence is embedded into a real-valued vector. Then, the sequence is passed through the encoder, which contains $M$ encoder blocks defined as follows:


\begin{align} \bar{h}_{m}=h_{m-1}+\mathrm{MHA}\left(\operatorname{LN}\left(h_{m-1}\right)\right) \end{align} 

\begin{align} h_{m}=\bar{h}_{m}+\operatorname{FFN}\left(\operatorname{LN}\left(\bar{h}_{m}\right)\right)\end{align}
where MHA is the multi-head attention layer, LN is layer-normalization \cite{ba2016layer}, and FFN is a feed-forward network. Crucially, the layers in each block sum its outputs to the outputs from the previous layers ($h_{m-1}$). This kind of connection is known as residual connection and it eases learning, enabling the training of deeper models \cite{DBLP:journals/corr/HeZRS15}. Regarding the MHA, it is defined as follows:


\begin{align} MHA = \text{softmax}\left(\frac{QK^T}{\sqrt{d_k}}\right)V\end{align}

Where $Q$ is the query matrix, $K$ is the key matrix, $V$ is the value matrix, and $d_k$ is the dimension of the keys. In encoder's (and decoder's) self-attention, all key, value, and value vectors come from linear projections from the input sequence (from the previous layer). Note that MHA is defined with matrices, and not vectors, because the implementation is batched.

Similarly, the input tokens from the target sequence are also embedded, and passed through the decoder attention-based blocks. There are two relevant differences with respect to the encoder. The first one is that decoder blocks have an additional MHA layer (followed by an additional LN) to perform encoder-decoder attention (as opposed to self-attention). The second one is that in training, causal masking is applied so that future tokens don't leak into the input of the decoder. Finally, the output embeddings from the decoder are projected back into the vocabulary space with an additional linear layer, so that, after applying the softmax function, the output is a probability distribution for the tokens in the vocabulary.

\subsection{Model Architecture and Training Details}
We use the modifications from BART~\cite{DBLP:journals/corr/abs-1910-13461}. Our model has 6 encoder layers, 6 decoder layers, an embedding size of 1024, a context window of 1024 positions for both source and target sequences, and shared embeddings for the encoder, decoder and decoder output layer. We initialize the parameters from $\mathcal{N}(0, 0.02)$. Instead of vanilla stochastic gradient descent, we use the Adam optimizer \cite{kingma2017adam}. Each model is trained for 72 hours on 4 Nvidia A100s. In total, we train 4 models (one for each evaluated architecture and optimization level).


We do not use  dropout \cite{JMLR:v15:srivastava14a} regularization method typically used in Transformer models, following the intuition, confirmed in preliminary experiments, that it would be detrimental to our task. Instead, we regularize with weight decay.

%% file: inference.tex
\section{Inference}
Figure~\ref{fig:DecompilerDiagram} shows the process
\name{} uses to decompile source code.
First, we use our trained model to generate
source code (section~\ref{sec:ModelInference}).
This generates a hypothesis C file. We then 
use type inference (Section~\ref{sec:TypeInferenceAlgorithm})
to infer the types that should be inserted.

\subsection{Model Inference}
\label{sec:ModelInference}
\name{} uses a trained model $T$
that takes as input some assembly $s$
and generates a C code hypothesis, $\hat{c}$. In the simplest case, with \textit{greedy} decoding, we would feed $s$ into the encoder, which would encode it in parallel. Then, autoregressively, we would keep taking the token that maximizes the probability, until predicting the special token EOS marking the end of the sequence. Since our goal is to maximize the global probability of the predicted sequence as opposed to the local probability of just the next token, we use beam search decoding with a beam size of $k=5$. That is, at each step, we keep the top $k$ hypotheses with the highest probability, and at the end of the decoding we select the first one passing the IO tests (if any). To speed up evaluation, we batch the examples on an NVIDIA A6000 GPU. However, the model is small enough to be run on consumer CPUs with decent latency.

\subsection{Type Inference}
\label{sec:TypeInferenceAlgorithm}

The C
code that our model outputs can feature missing
types.  For example, it may be that $T$
generates code that relies on some
type \texttt{my\_int} that it has
frequently seen in training, but is not part of the
standard library.  We use type inference (Section~\ref{sec:TypeInferenceAlgorithm})
to solve this problem. For any of these types
we use PsycheC~\cite{Melo2018} which is a tool that infers types
that respect C's type system.
PsycheC takes a partial program, $\mathcal{P}$
as its input and produces
a complete program $\mathcal{P'}$ that
can be compiled.

It performs three key steps to do this.
First, PsycheC parses partial C programs,
which can contain ambiguities (e.g.,
\texttt{(a)*b} could be a cast or a unary
operation or a binary operation depending
on whether \texttt{a} is a variable or a type).

After parsing, PsycheC
produces a series
of constraints that must be applied
to generated types.
For example, if we have some type
$\tau$ that is used in an assignment
\texttt{$\tau$ a = b;}, and we know \texttt{b}
is of type $\tau'$, we can conclude that
$\tau = \tau'$.  PsycheC builds a set of rules
for type equivalence e.g., $\tau_1 = \tau_2 \implies
\tau_1* = \tau_2*$ and uses syntax-directed
generation of constraints on types within
this framework.  For example,
\texttt{*E} has type $*\tau$ if \texttt{E}
has type $\tau$.  To deal with
the ambiguities discussed above,
PsycheC uses a \textit{lattice} model
of constraint generation.

Once these constraints are generated,
PsycheC solves them and produces types that
make the program compile.
We inject these generated dependencies into
our generated decompiled code, checking that there is no conflict with the previously defined code. Once we have compilable code, we test it using IO examples
as discussed in Section~\ref{sec:IOCorrectness}.

%% file: setup.tex
\section{Results}
\label{sec:results}
We evaluate \name{} against alternative approaches 
on two different ISAs, at two different optimization levels on two
different benchmark suites.

\subsection{Experimental Setup}

\subsubsection{Benchmarks}
We evaluate our approach on two benchmark suites: the small-scale 112 program synthesis benchmark used in~\cite{bruce_synth}, Synth, to examine in detail decompilation behavior;  and a subset of ExeBench \cite{exebench} to allow wider-scale evaluation and interaction with user-defined types and external function calls, a challenging task for decompilation. For all models and baselines, we discard the benchmarks in which GCC couldn't compile the original C code in our host machine used for evaluation. For a more realistic evaluation, we do not discard any benchmark based on length. None of the about 4000 test benchmark programs were seen during training as per the performed token-level hash-based deduplication. We report the normalized edit similarity with respect to the ground truth, and the percentage of functions that pass the IO tests (Section~\ref{sec:correct}). That is: how many decompiled programs when recompiled,  give the same
output as the original assembly for different inputs?
We evaluate  on two different ISAs, x86 and ARM.
We consider two levels of code optimization -O0 and -O3 to evaluate 
the impact of code complexity.

\begin{figure*}[h!]
\begin{tabular}{cc}
\includegraphics[width=0.47\textwidth]{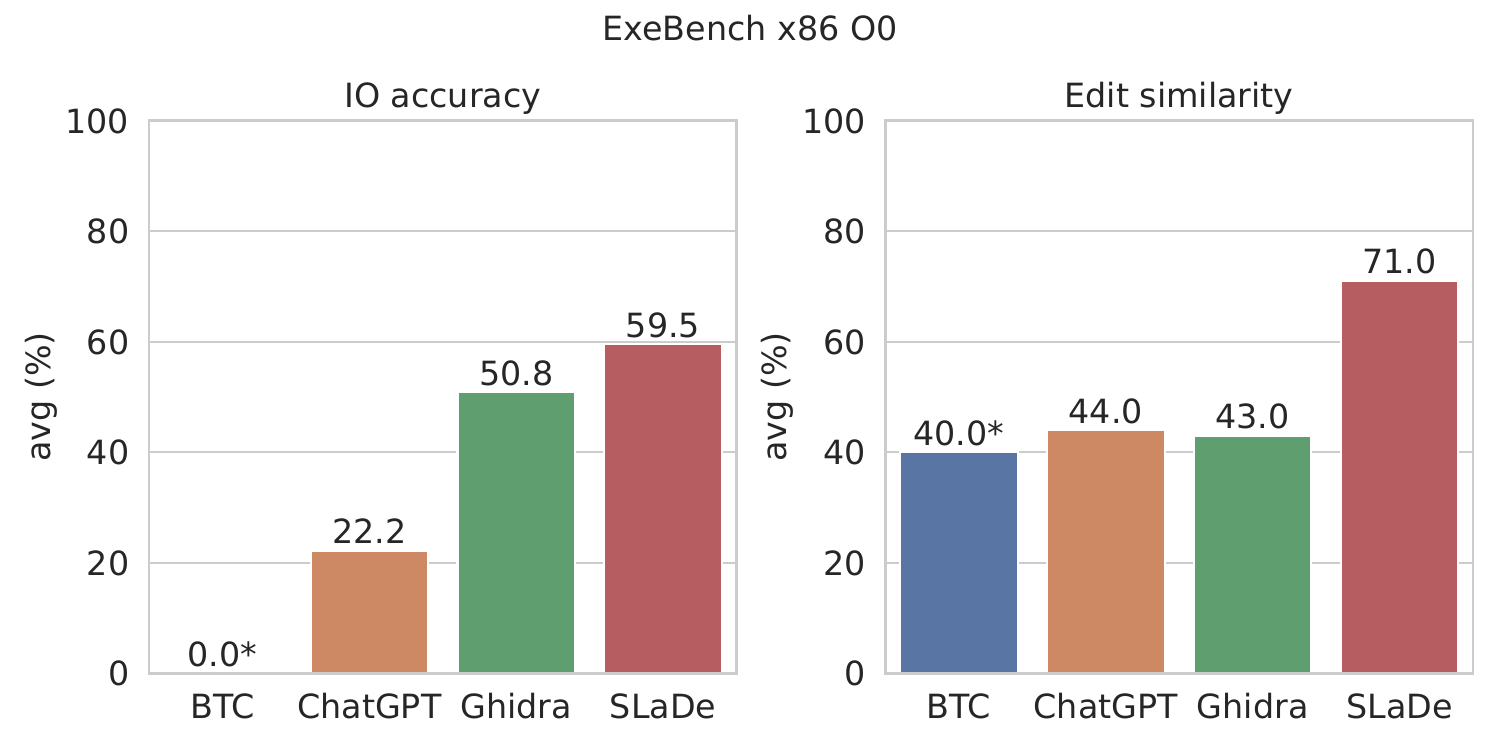}
 &
\includegraphics[width=0.47\textwidth]{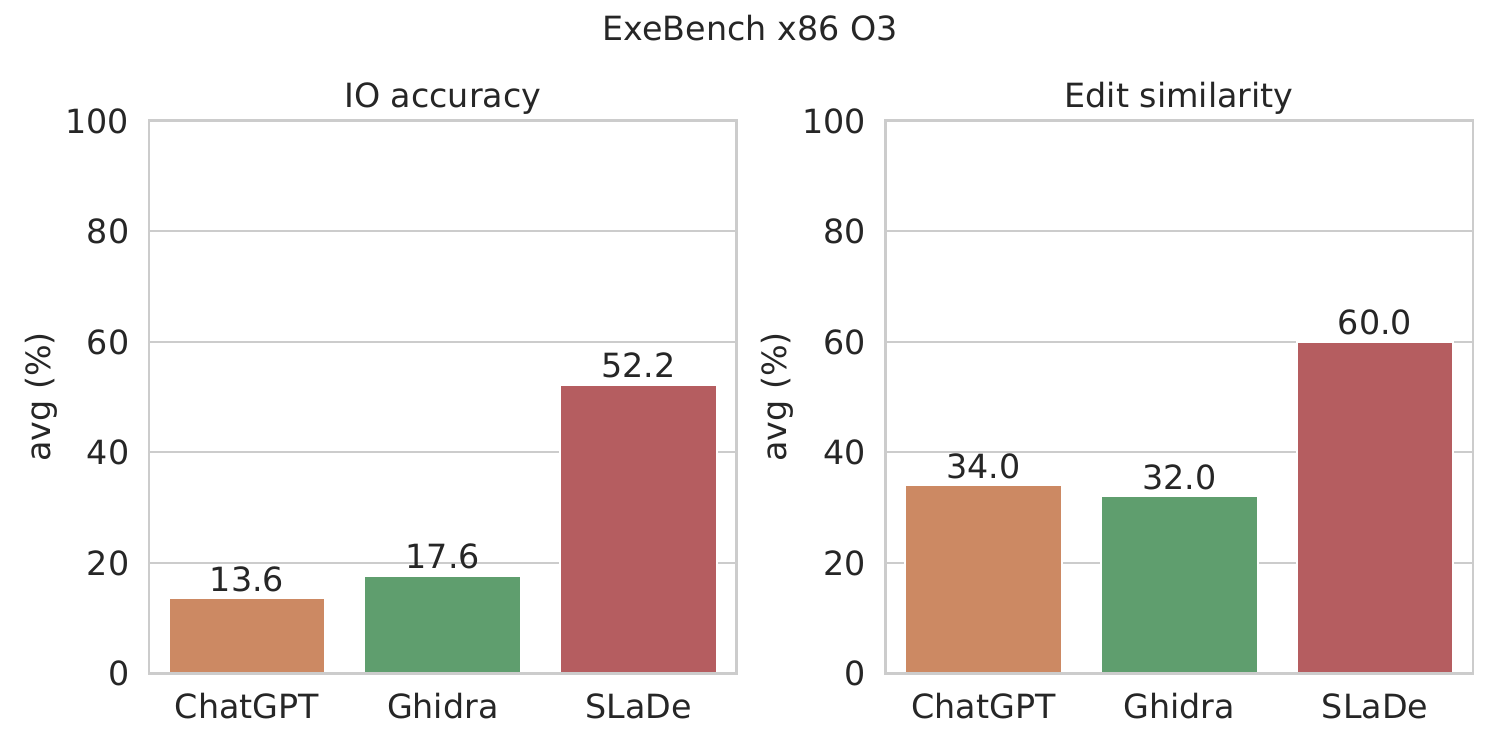}
\end{tabular}
\label{sec:exp}
\caption{ExeBench, x86: -O0 (left)  -O3 (right),  input-output (IO) accuracy and edit similarity. 
A decompiled program is IO accurate if it gives the same outputs for the same range of inputs as the original assembly. 
BTC's edit distance is as-repored in~\cite{Hosseini2022} on a different dataset. Its dataset is restricted to -O0 and does
not support evaluations of correctness, so omitted.  \name{}
out-performs existing techniques, producing 1.17x to 3.83x 
more correct code, and a higher edit similarity
than Ghidra, ChatGPT and BTC.
. }
\label{fig:diagram_A_x}
\end{figure*}
\begin{figure*}[h!]
\begin{tabular}{cc}
\includegraphics[width=0.47\textwidth]{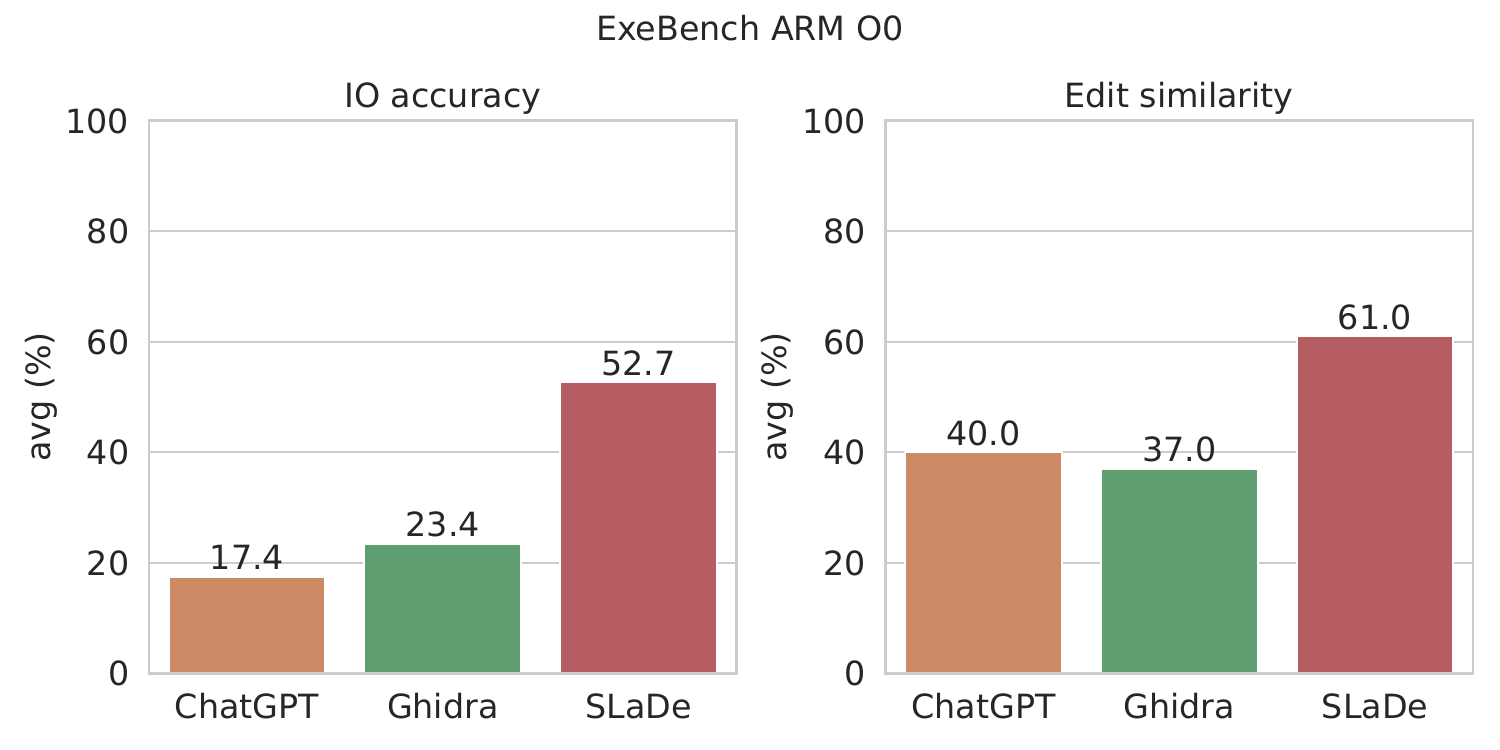}
 &
\includegraphics[width=0.47\textwidth]{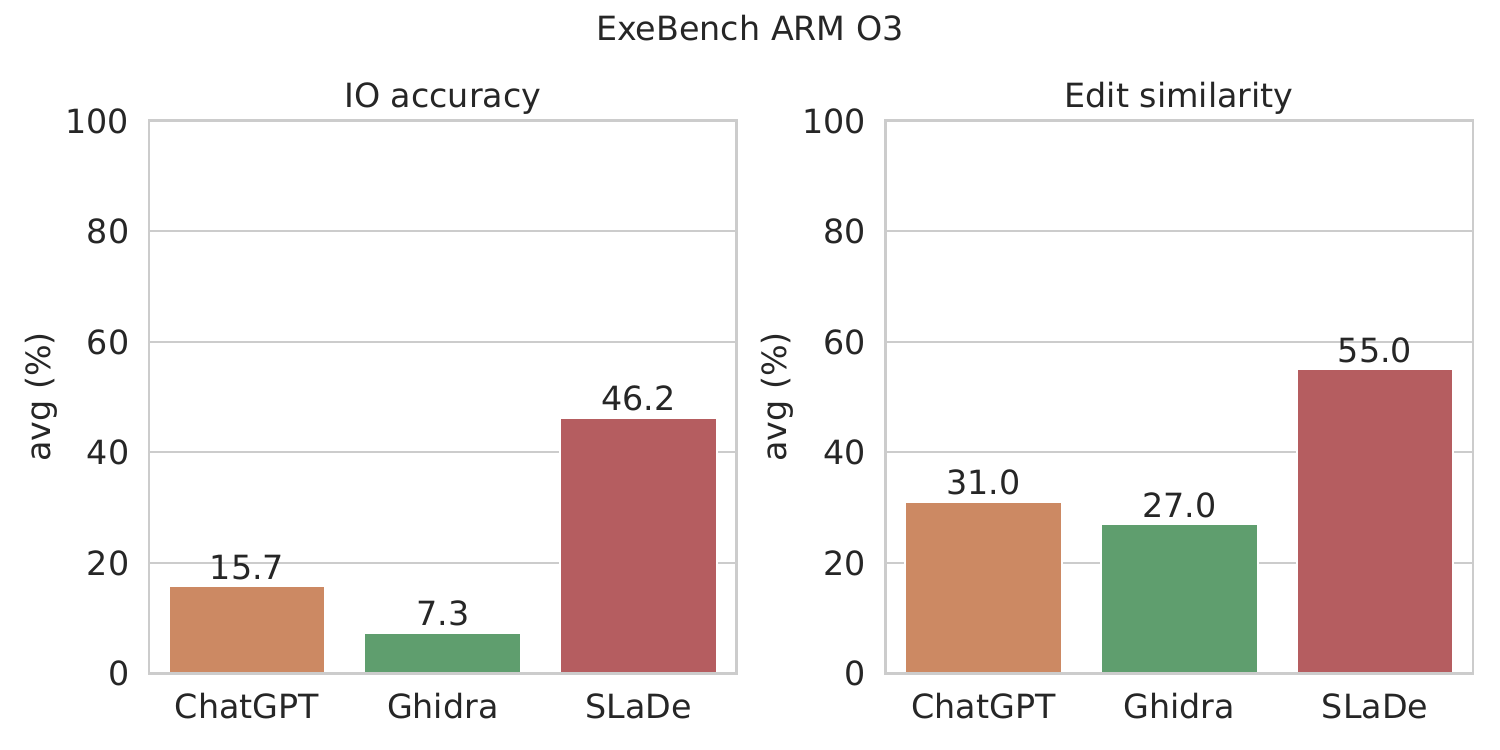}
\end{tabular}
\caption{ExeBench, x86: -O0 (left)  -O3(right),  IO accuracy and edit similarity.
\name{} significantly 
out-performs existing techniques producing 2.2x to 6.32x more accurate code and a higher edit similarity.}
\label{fig:diagram_A_A}
\end{figure*}

\begin{figure*}[h!]
\begin{tabular}{cc}
\includegraphics[width=0.47\textwidth]{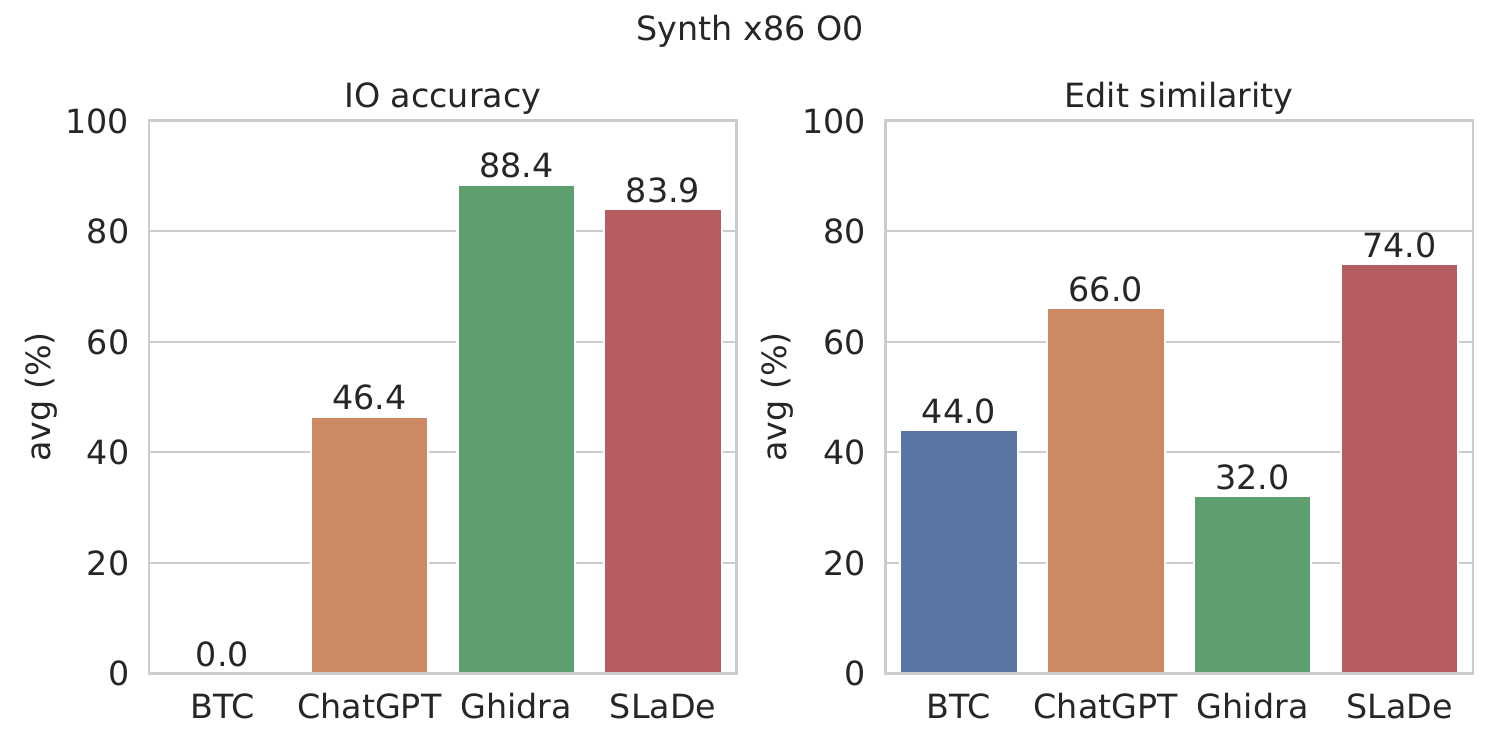}
 &
\includegraphics[width=0.47\textwidth]{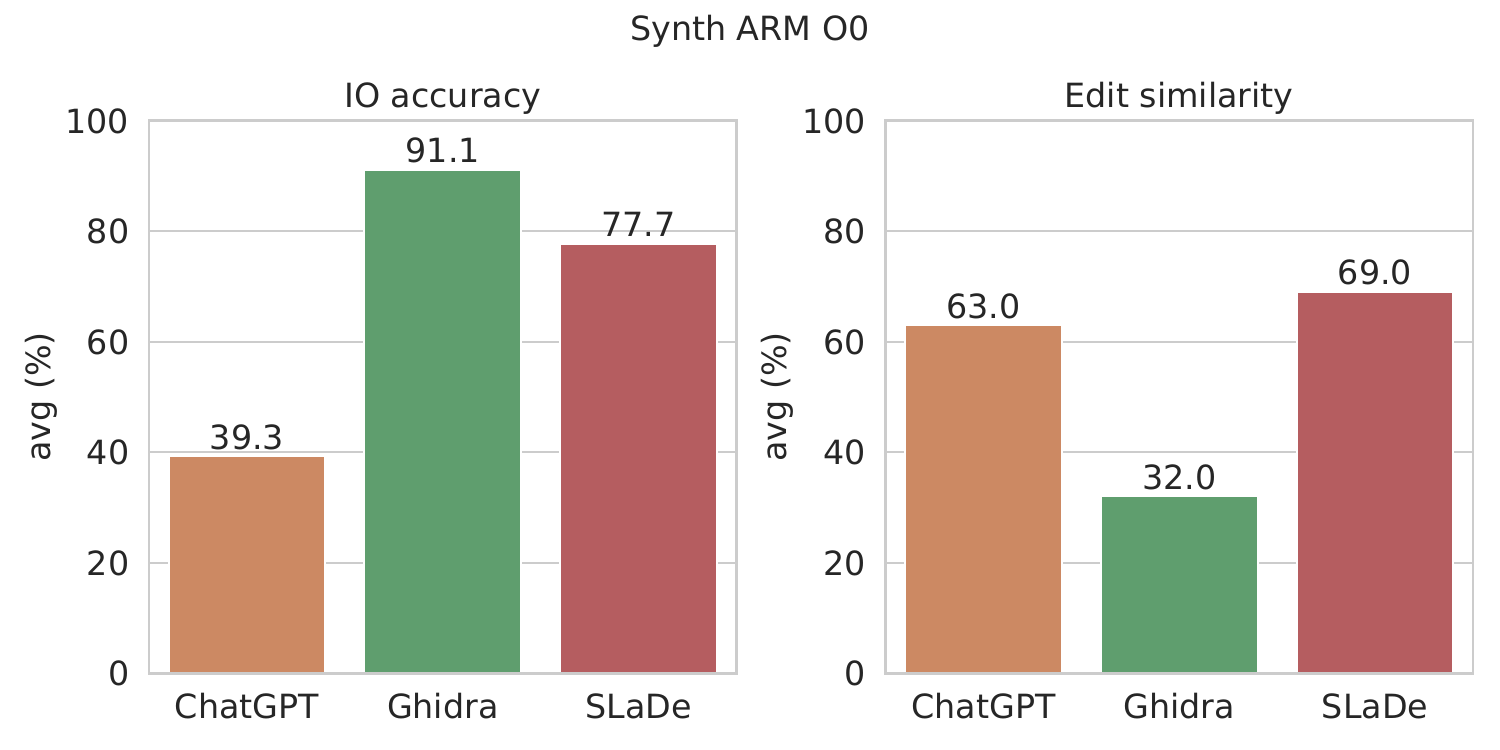}
\end{tabular}
\caption{Synth -O0: x86 (left) and ARM (right), IO accuracy  and edit similarity. BTC edit similarity is experimentally evaluated. Ghidra has 
slightly higher IO accuracy than \name{} on these simpler, unoptimized, benchmarks. }
\label{fig:diagram_S_0}
\end{figure*}

\begin{figure*}[h!]
\begin{tabular}{cc}
\includegraphics[width=0.47\textwidth]{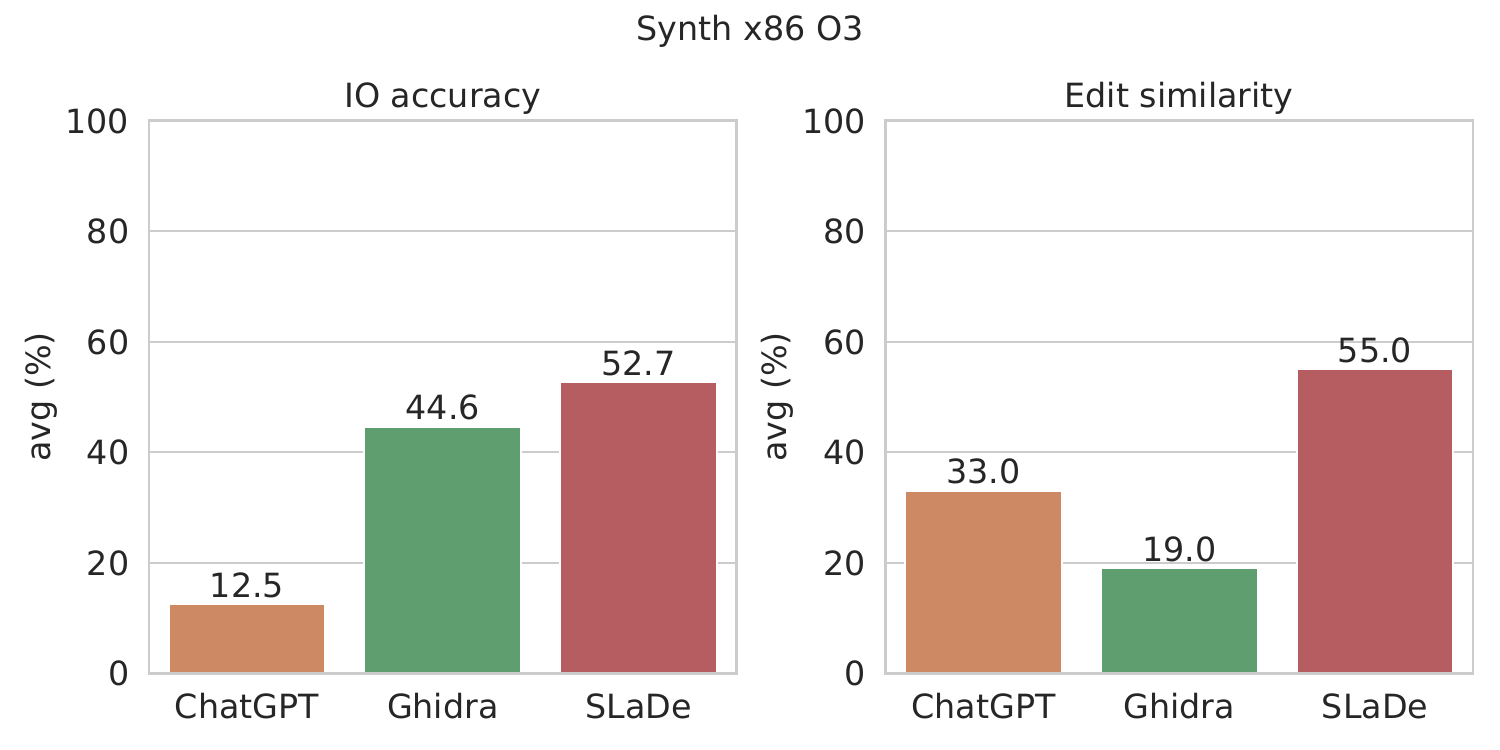}
 &
\includegraphics[width=0.47\textwidth]{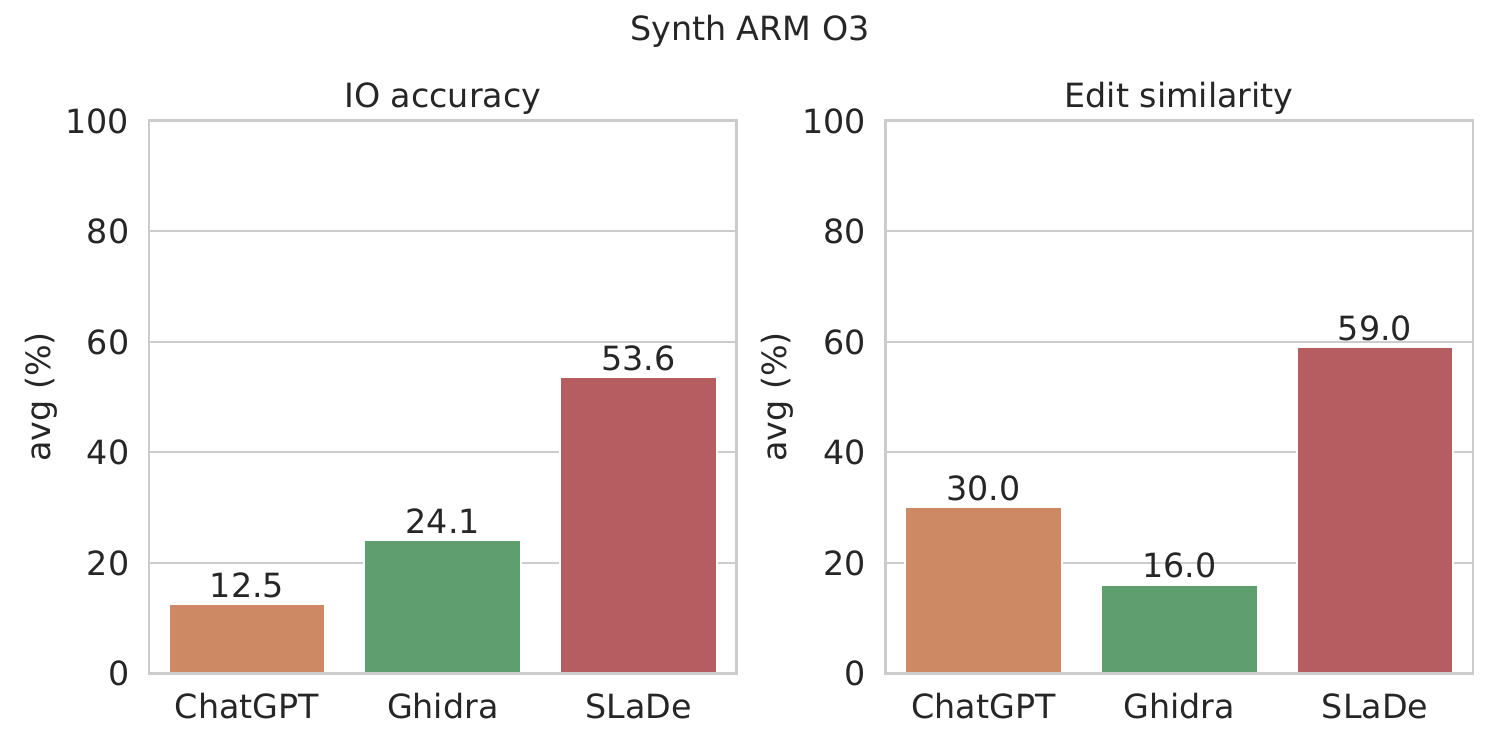}
\end{tabular}
\caption{Synth -O3: x86 and ARM, IO accuracy  and edit similarity. \name{} has a small reduction in accuracy compared to O0 while Ghidra is more negatively affected. \name{} produces 1.81x to 4.28x more accurate code.}
\label{fig:diagram_S_3}
\end{figure*}

\begin{figure}
\includegraphics[width=0.40\textwidth]{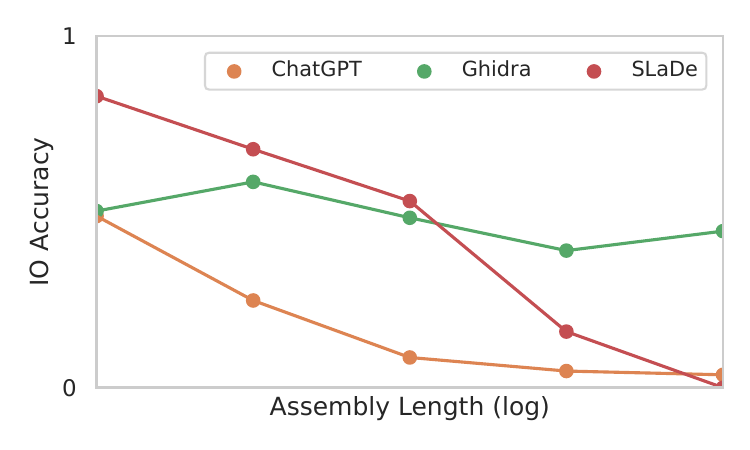}
\caption{
IO accuracy as the program size changes.
}
\label{fig:IOCorrectnessVsLength}
\end{figure}
\begin{figure}
\includegraphics[width=0.40\textwidth]{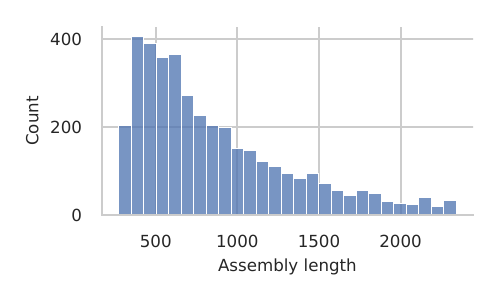}
\caption{
Distribution of program lengths
in ExeBench by character length. 
}
\label{fig:AssemblyLengthDistribution}
\end{figure}

\begin{figure}
    \centering
 \includegraphics[width=0.45\textwidth]{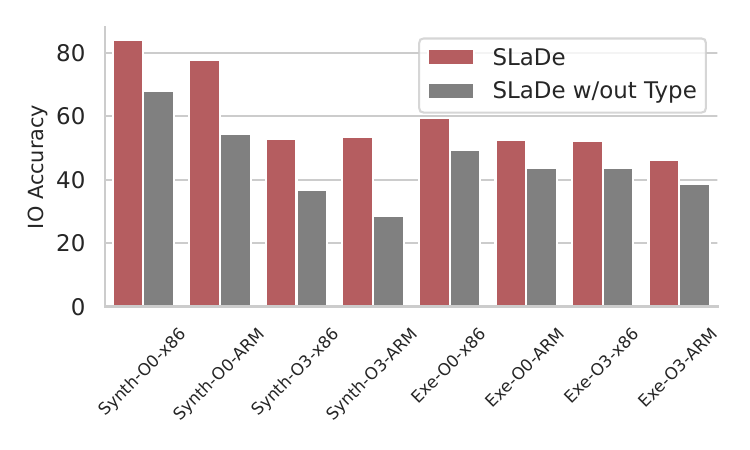}
    \caption{\name{} with and
    without type-inference (Section~\ref{sec:TypeInferenceAlgorithm}).  On average,
    the type inference algorithm makes \name{} 14\%
    better.}
    \label{fig:TypeInferenceGraph}
\end{figure}

\subsubsection{Decompilers}


\paragraph{Ghidra} 
We invoke the headless version of the Ghidra decompiler, providing the object file of the compiled function assembly generated by GCC.
%
We then insert the resulting C code function into the original calling program which contains context such as  type declarations.  
Ghidra frequently makes reference to standard types that may not be declared in the original 
program such as {\tt bool}. For Ghidra alone, we  insert those types into the header of the program for fair evaluation.

\paragraph{ChatGPT} is a well known chatbot.
with a wide range of uses, including program generation. 
We used a best-effort, 
trial-and-error approach to find an effective prompt.  The following format gave the best response:
\texttt{
Decompile the following ARM assembly function into C code. Return the code and only the code, no explanations or introductions.}

We also experimented with \textit{few-shot learning}, that is, including examples of the intended decompilation task in the prompt itself, but this harmed rather than helped due to the very long sequences needed to include C-assembly pairs into the prompt.  
We inserted the resulting function into
the original calling program as for Ghidra.
\paragraph{BTC} is the only publicly available alternative neural-decompiler 
that tackles real-world assembly.
We run it on Synth  benchmark using the model provided by the authors. Elsewhere, we use the  value reported in their paper (on another dataset), in the case of edit similarity, or the accuracy from the model limitations stated in the paper.
\paragraph{\name{}}
After invoking \name{} we insert the resulting function into
the original calling program as for Ghidra.
\name{} is implemented  
in PyTorch~\cite{paszke2017automatic} using Fairseq~\cite{ott-etal-2019-fairseq} and Huggingface Transformers libraries~\cite{wolf-etal-2020-transformers}. 

%% file: results.tex
\input{final_plots/correlations_AnghaBench}

\subsection{x86 Decompilation}
\label{sec:EditDistanceResult}

Figure \ref{fig:diagram_A_x} shows the performance of \name{} relative to the alternative  techniques.
The two leftmost  graphs
show IO accuracy and  edit similarity on  unoptimized (-O0), x86 code from ExeBench. The rightmost two show the same results for optimized (-O3), x86 code.

\paragraph{Unoptimized -O0 decompilation}

\name{} is able to accurately decompile almost 60\% of the holdout ExeBench
test suite, 17\% more  than
Ghidra and more than double ChatGPT. BTC is unable to accurately
decompile any of the programs accurately which is not surprising given its 
limited focus. 
\name{}'s failures are largely due to incorrectly predicting the arguments to externally   declared functions. Ghidra has the
same issue as well as additional type errors due to structs.

\name{}  is able to produce highly readable code with an edit similarity of over 70\%. This was a significant improvement over the other approaches:  40\% to 44\% similarity.  The value for BTC
is a reported values from \cite{Hosseini2022}  and is close to that experimental value of
ChatGPT. In section \ref{sec:synth} we experimentally evaluate it on the smaller Synth bench. Surprisingly Ghidra has a similar edit similarity to ChatGPT.


\paragraph{Optimized -O3  decompilation}

While {\tt SLaDe}  is able to achieve impressive 
performance on unoptimized code, in practice most shipped code  is  
optimized and the most likely use case for decompilation.
The two right-most graphs show x86, -O3 results. We do not include BTC
as it targets unoptimized (-O0) code.
All approaches unsurprisingly find -O3 more challenging. However the relative 
decline in IO accuracy for {\tt SLaDe} is less than for the other schemes.
{\tt SLaDe} is now able to deliver almost three times  the number of correctly decompiled programs as Ghidra and almost four times  that of  ChatGPT. Furthermore, the readability of the code generated by {\tt SLaDe}  is nearly twice that
of ChatGPT and Ghidra. 


\subsection{ARM Decompilation}
In principle one of the main benefits of neural decompilation
is that it can be straightforwardly trained and applied to new ISAs with no re-engineering effort.
Figure \ref{fig:diagram_A_A} shows  performance on a new ISA, ARM,
with the two leftmost  graphs
showing IO accuracy and  edit similarity for unoptimized (-O0)  code from ExeBench. The rightmost two graphs show the same results for optimized (-O3) code.
Compared to x86, \name{} has a slight degradation in accuracy and edit similarity. However, Ghidra's accuracy  degrades  by at least a  factor of two. \name{} is now more than $2\times$ as accurate on -O0 and more than $6\times$ more accurate at -O3. ARM's stack calling conventions appear to be  an issue  
for Ghidra. ChatGPT is less affected
though still 3$\times$ less accurate than \name{}.

\subsection{Discussion}
Despite being rule-based, Ghidra
has a key restriction.
In cases of external type/function declarations, unlike \name{}, 
Ghidra
does not generate types but 
rather leaves them undefined. 
 This
accounts for many 
of the IO correctness failures (due to lack of compilability) it
suffers.  While a programmer could resolve these, given
the challenges in understanding
Ghidra's code, filling in the missing
dependencies is a non-trivial task.\footnote{\url{https://github.com/NationalSecurityAgency/ghidra/issues/236}} In the next section, we therefore
evaluate using simpler programs where type ambiguity
is not an issue.

\begin{figure*}
\begin{tabular}{cc}
\includegraphics[width=0.45\textwidth]{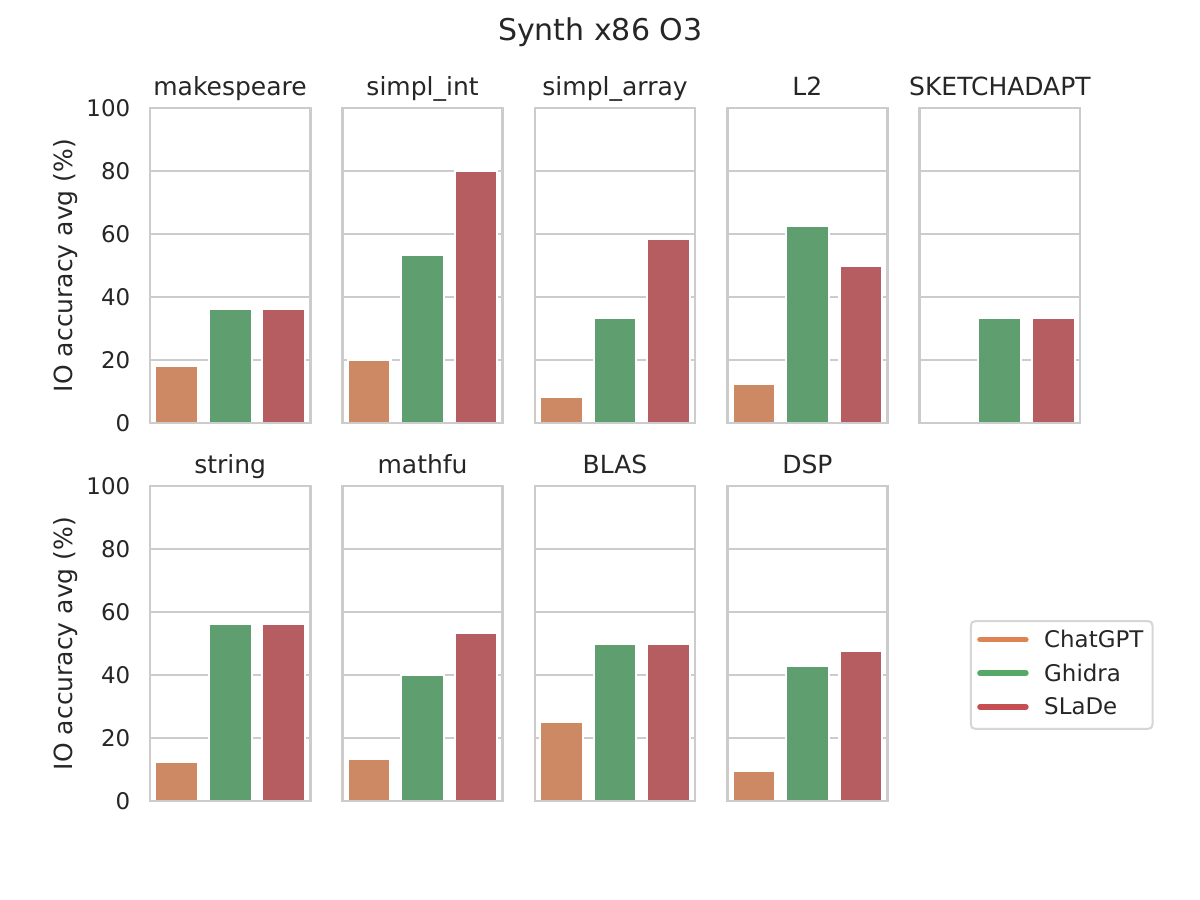} & 
\includegraphics[width=0.45\textwidth]{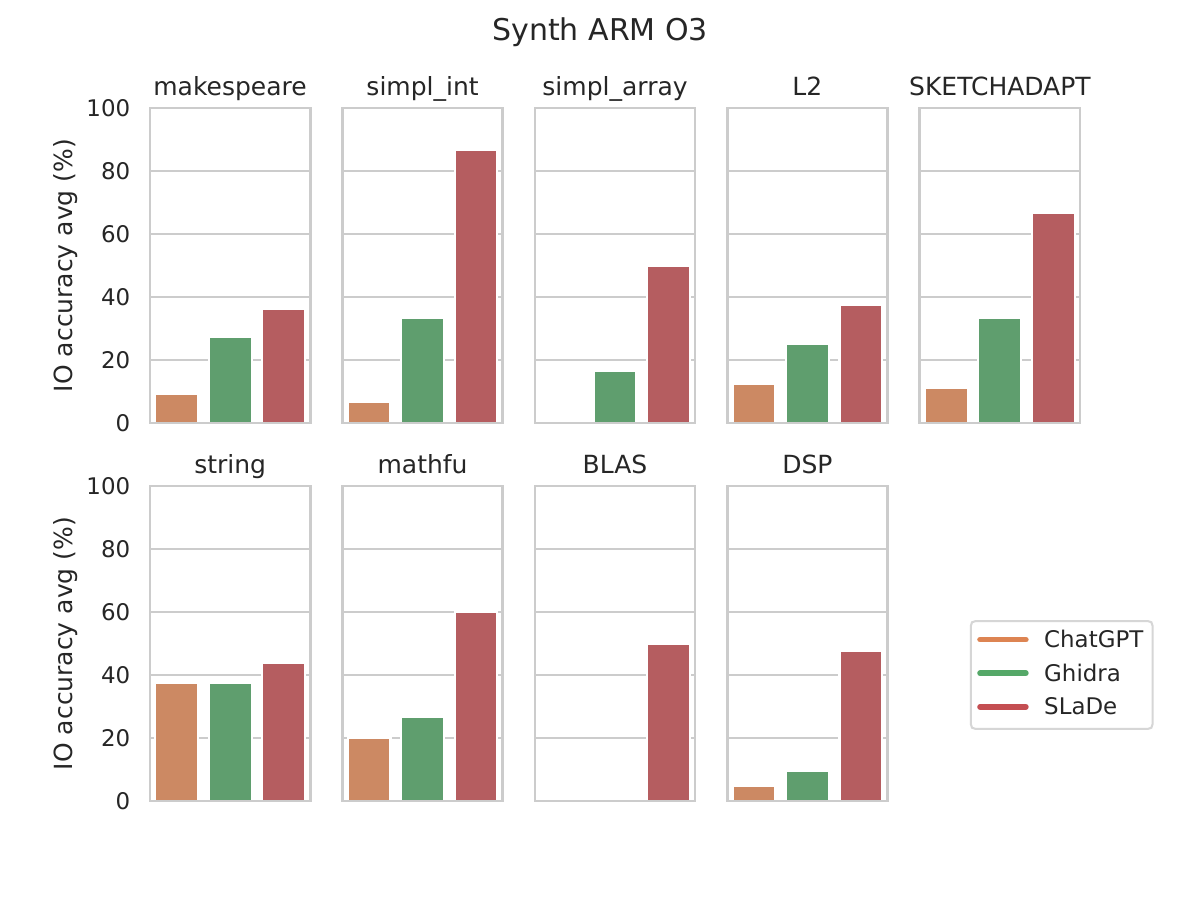}
\end{tabular}
\caption{Analysis of IO accuracy depending on program type and optimization level.  We see that
\name{} performs significantly better
on simple integer benchmarks in simpl\_int, achieving more than 80\% in both cases
than it does on the functional programming-based benchmarks in \textsc{SketchAdapt} where it achieves 40\% and 60\% on x86 and Arm respectively.}
\label{fig:groups}
\end{figure*}

\subsection{Synthesis Benchmark}
\label{sec:synth}
Here we evaluate on a set of  benchmarks with simpler types.
Figure \ref{fig:diagram_S_0} shows the performance of {\tt SLaDe} against  alternative approaches.

\paragraph{Unoptimized x86, ARM}
On unoptimized assembly, both {\tt SLaDe} and  Ghidra  perform well with over 80\% IO accuracy, showing that these benchmarks are easier to decompile. They contain only simple
argument types and do not call external functions.
Ghidra outperforms {\tt SLaDe} by 5\% on x86 
and 20\% on ARM but is significantly worse in the readability of the generated program. Surprisingly Ghidra has worse readability on these benchmarks relative to ExeBench. {\tt SLaDe} has  more than double the edit similarity of Ghidra
for both ISAs. 
ChatGPT is significantly less accurate than both {\tt SLaDe} and Ghidra. It has better readability than Ghidra but is inferior
to {\tt SLaDe} on both ISAs. We also evaluated BTC on this smaller set of benchmarks where it was shown to have a better edit similarity 44\% than its reported
result 40\%. 

\paragraph{Optimized x86, ARM}
Figure \ref{fig:diagram_S_3} again shows that
optimization  has a dramatic effect on Ghidra, particularly on the ARM ISA
where IO accuracy has reduced by over two-thirds. The performance of ChatGPT similarly falls. Complex types
are no longer the problem for Ghidra, rather the obfuscating effect of compiler optimization has degraded its performance.
In the optimized setting 
{\tt SLaDe}  has superior accuracy for both ISAs and  superior readability. While
the readability of Ghidra and ChatGPT almost half when optimizing, {\tt SLaDe} 
only suffers a small degradation.

\subsection{\name{} failure}
While \name{} performs well, here we cover two
typical failure cases. For the ground truth
code:
\begin{lstlisting}[language=C,basicstyle=\footnotesize]
void satd8x8_getDiff(unsigned int res[]) {
  memcpy(res, &mat[lastRow * 8],
        8 * sizeof(*mat));
  lastRow++;
}
\end{lstlisting}
\name{} produces:
\begin{lstlisting}[language=C,basicstyle=\footnotesize]
void satd8x8_getDiff( float *mat ) {
  memcpy( mat, &mat[lastRow*8], 32);
  lastRow++;
}
\end{lstlisting}
Critically, \name{} mistypes the argument
to the function, as most matrix variables it
encounters are of \texttt{float} type.
Similarly, for the ground truth
code:
\begin{lstlisting}[language=C,basicstyle=\footnotesize]
void clock_add( SClock *clk, double incr ) {
  if(clk) {
    clk->curtime += incr;
    clk->basetime += incr;
    clk->seqno++;
  }
}
\end{lstlisting}
\name{} produces:
\begin{lstlisting}[language=C,basicstyle=\footnotesize]
void clock_add(struct clock *__restrict ev,
               double d) {
  if(ev) {
    ev->constev += d;
    ev->constsp++;
    ev->constt--;
  }
}
\end{lstlisting}
Which is the right idea, but aside from different struct and field names, it uses the \texttt{++} and the \texttt{-}\texttt{-}
operations rather than the \texttt{+=incr} and \texttt{++} operations, respectively.
When the original code was compiled with -O0,
Ghidra produced correct decompiled code.  However, with -O3
GCC's optimizations obscuring the resulting assembly, Ghidra was not able to produce compilable code.

%% file: final_plots/correlations_AnghaBench.tex
\begin{table*}
\caption{Pearson's correlation coefficient between code features and IO accuracy on ExeBench.}

\begin{tabular}{lrrr|rrr|rrr|rrr}
 & \multicolumn{3}{c}{x86 O3} & \multicolumn{3}{c}{x86 O0} & \multicolumn{3}{c}{ARM O0} & \multicolumn{3}{c}{ARM O3} \\
 & ChatGPT & Ghidra & \name{} & ChatGPT & Ghidra & \name{} & ChatGPT & Ghidra & \name{}& ChatGPT & Ghidra &\name{} \\
Compiles & \cellcolor[HTML]{f29072} \color[HTML]{f1f1f1}  0.54 & \cellcolor[HTML]{d44e41} \color[HTML]{f1f1f1}  0.81 & \cellcolor[HTML]{d65244} \color[HTML]{f1f1f1}  0.80 & \cellcolor[HTML]{f29274} \color[HTML]{f1f1f1}  0.53 & \cellcolor[HTML]{d55042} \color[HTML]{f1f1f1}  0.81 & \cellcolor[HTML]{c0282f} \color[HTML]{f1f1f1}  0.94 & \cellcolor[HTML]{f49a7b} \color[HTML]{000000}  0.49 & \cellcolor[HTML]{ba162b} \color[HTML]{f1f1f1}  0.96 & \cellcolor[HTML]{bd1f2d} \color[HTML]{f1f1f1}  0.95 & \cellcolor[HTML]{ee8669} \color[HTML]{f1f1f1}  0.59 & \cellcolor[HTML]{b70d28} \color[HTML]{f1f1f1}  0.98 & \cellcolor[HTML]{d85646} \color[HTML]{f1f1f1}  0.78 \\
Edit Similarity & \cellcolor[HTML]{f7ba9f} \color[HTML]{000000}  0.32 & \cellcolor[HTML]{efcfbf} \color[HTML]{000000}  0.16 & \cellcolor[HTML]{ee8669} \color[HTML]{f1f1f1}  0.59 & \cellcolor[HTML]{f4c6af} \color[HTML]{000000}  0.24 & \cellcolor[HTML]{dddcdc} \color[HTML]{000000}  0.01 & \cellcolor[HTML]{f39475} \color[HTML]{000000}  0.53 & \cellcolor[HTML]{f1ccb8} \color[HTML]{000000}  0.19 & \cellcolor[HTML]{e6d7cf} \color[HTML]{000000}  0.07 & \cellcolor[HTML]{ef886b} \color[HTML]{f1f1f1}  0.58 & \cellcolor[HTML]{f4c5ad} \color[HTML]{000000}  0.25 & \cellcolor[HTML]{ecd3c5} \color[HTML]{000000}  0.13 & \cellcolor[HTML]{ee8468} \color[HTML]{f1f1f1}  0.60 \\
ASM Length& \cellcolor[HTML]{cfdaea} \color[HTML]{000000}  -0.10 & \cellcolor[HTML]{e0dbd8} \color[HTML]{000000}  0.03 & \cellcolor[HTML]{cbd8ee} \color[HTML]{000000}  -0.13 & \cellcolor[HTML]{c4d5f3} \color[HTML]{000000}  -0.18 & \cellcolor[HTML]{d5dbe5} \color[HTML]{000000}  -0.06 & \cellcolor[HTML]{afcafc} \color[HTML]{000000}  -0.30 & \cellcolor[HTML]{c0d4f5} \color[HTML]{000000}  -0.20 & \cellcolor[HTML]{cad8ef} \color[HTML]{000000}  -0.14 & \cellcolor[HTML]{a6c4fe} \color[HTML]{000000}  -0.36 & \cellcolor[HTML]{cbd8ee} \color[HTML]{000000}  -0.13 & \cellcolor[HTML]{d8dce2} \color[HTML]{000000}  -0.04 & \cellcolor[HTML]{c6d6f1} \color[HTML]{000000}  -0.16 \\
C Length & \cellcolor[HTML]{d3dbe7} \color[HTML]{000000}  -0.07 & \cellcolor[HTML]{dcdddd} \color[HTML]{000000}  -0.01 & \cellcolor[HTML]{cad8ef} \color[HTML]{000000}  -0.14 & \cellcolor[HTML]{cdd9ec} \color[HTML]{000000}  -0.11 & \cellcolor[HTML]{d8dce2} \color[HTML]{000000}  -0.04 & \cellcolor[HTML]{bfd3f6} \color[HTML]{000000}  -0.21 & \cellcolor[HTML]{cedaeb} \color[HTML]{000000}  -0.11 & \cellcolor[HTML]{d6dce4} \color[HTML]{000000}  -0.05 & \cellcolor[HTML]{bfd3f6} \color[HTML]{000000}  -0.21 & \cellcolor[HTML]{d1dae9} \color[HTML]{000000}  -0.09 & \cellcolor[HTML]{dadce0} \color[HTML]{000000}  -0.02 & \cellcolor[HTML]{c9d7f0} \color[HTML]{000000}  -0.15 \\
Num Func Args & \cellcolor[HTML]{d2dbe8} \color[HTML]{000000}  -0.08 & \cellcolor[HTML]{dcdddd} \color[HTML]{000000}  -0.01 & \cellcolor[HTML]{ccd9ed} \color[HTML]{000000}  -0.12 & \cellcolor[HTML]{d3dbe7} \color[HTML]{000000}  -0.07 & \cellcolor[HTML]{e6d7cf} \color[HTML]{000000}  0.07 & \cellcolor[HTML]{cbd8ee} \color[HTML]{000000}  -0.13 & \cellcolor[HTML]{d3dbe7} \color[HTML]{000000}  -0.07 & \cellcolor[HTML]{f2cbb7} \color[HTML]{000000}  0.20 & \cellcolor[HTML]{d1dae9} \color[HTML]{000000}  -0.09 & \cellcolor[HTML]{d2dbe8} \color[HTML]{000000}  -0.08 & \cellcolor[HTML]{e1dad6} \color[HTML]{000000}  0.04 & \cellcolor[HTML]{cfdaea} \color[HTML]{000000}  -0.10 \\
Num Pointers& \cellcolor[HTML]{d8dce2} \color[HTML]{000000}  -0.04 & \cellcolor[HTML]{e0dbd8} \color[HTML]{000000}  0.03 & \cellcolor[HTML]{c6d6f1} \color[HTML]{000000}  -0.16 & \cellcolor[HTML]{d8dce2} \color[HTML]{000000}  -0.03 & \cellcolor[HTML]{edd1c2} \color[HTML]{000000}  0.14 & \cellcolor[HTML]{c5d6f2} \color[HTML]{000000}  -0.17 & \cellcolor[HTML]{d8dce2} \color[HTML]{000000}  -0.04 & \cellcolor[HTML]{f2cbb7} \color[HTML]{000000}  0.20 & \cellcolor[HTML]{ccd9ed} \color[HTML]{000000}  -0.12 & \cellcolor[HTML]{d5dbe5} \color[HTML]{000000}  -0.06 & \cellcolor[HTML]{e7d7ce} \color[HTML]{000000}  0.08 & \cellcolor[HTML]{cad8ef} \color[HTML]{000000}  -0.14 \\
\end{tabular}
\label{tab:corr}

\end{table*}

%% file: analysis.tex
\section{Analysis}
\label{sec:analysis}
In this section, we analyze the results 
and explore the limitations of traditional and  
language-model-based
decompilation.

\subsection{Impact of Assembly Length on Accuracy }
Figure \ref{fig:IOCorrectnessVsLength} shows how IO accuracy similarity varies for increasing assembler
length for ExeBench, x86 -O0.  All three approaches are better at decompiling shorter assembler sequences, since increased assembly complexity makes decompilation more challenging. Both ChatGPT and {\tt SLaDe} have a steeper decline than Ghidra, illustrating that neural schemes are more sensitive to the distribution type of training examples than hand-crafted approaches.  Figure~\ref{fig:AssemblyLengthDistribution} shows the distribution of assembly length in ExeBench.
We see that there is indeed  a bias to shorter-length assembly.




\subsection{Impact of type inference}

Figure~\ref{fig:TypeInferenceGraph} ablates the impact of type inference (Section~\ref{sec:TypeInferenceAlgorithm}).  On average, using type inference improves \name{} by $14\%$ by increasing the number of
decompiled examples that can compile.
These $14\%$ are the samples
for which \name{} generates
correct code, but was missing the required
typedefs.  The type inference algorithm was 
then able to infer the correct typedefs to make the code execute
correctly.

\subsection{Features}
To understand what factors affect each of the three decompilation schemes, we examined the correlation of IO
accuracy to different characteristics or features of the 
assembler programs. The features considered are: whether the code compiles, the edit similarity, the assembler 
length, the ground truth C program length, the number of function arguments and the number of pointers. Table \ref{tab:corr} shows 
the correlation coefficient for each approach across
each ISA and optimization level. 

Not surprisingly, whether a program compiles is important across all approaches and ISAs. A program that does not compile is unlikely to be IO accurate. However, this correlation is less significant  for ChatGPT which is able to hallucinate programs that do compile but are incorrect. Edit similarity is important for the neural schemes, particularly \name{}, but less so for Ghidra 
which is less concerned with readability. All other features are more weakly correlated. There is a small negative correlation with assembler and C length suggesting longer programs are harder to decompile. 
The number of functions and pointers is negatively correlated  for ChatGPT and \name{} but weakly positive for Ghidra. However, the magnitude is small and no conclusions can be 
drawn.

\subsection{Impact of Program Type}

The Synth benchmarks are broken down into groups.
In Figure \ref{fig:groups}, for x86, the easiest benchmarks to decompile were the simple integer
ones, Simple\_int, which  consists of
integer types and arithmetic with trivial control-flow
 Sketchadapt programs were the hardest. consisting of more complex string manipulation programs. ChatGPT performed best on BLAS problems, presumably from their  availability from web crawlings,  but generally performed poorly across groups, failing on all Sketchadapt problems. In four groups Ghidra and {\tt SLaDe}
have equal accuracy. In one case, Ghidra performs better on the L2 group which consists
of functional problems while {\tt SLaDe} performs better on the remaining three.

On Arm, different behavior is seen with {\tt SLaDe} outperforming across all categories. ChatGPT is able to decompile a Sketchapdapt problem but is now unable to manage any BLAS problems, its best-performing category on x86. Ghidra is also unable to decompile any of this group

\


%% file: related.tex
\section{Related Work}
\label{sec:rel}

\paragraph{Analytic decompilation}

Traditional approaches based on program analysis~\cite{retdec,ghidra}
represent years of hand-coded effort~\cite{Katz2019}
and rely on large bodies of pattern-matching
rules~\cite{hex}.  Research
work in this direction has focused
on correctness, through studies~\cite{liu2020far} and
formal methods~\cite{dasgupta2020scalable}.
LLVM IR is a popular lifting target,
to enable retargeting to new ISAs~\cite{yadavalli2019raising,Anand2010}.
More recent work lifts LLVM IR to OpenMP C
to exploit parallelism while retaining meaningful names~\cite{tan2023splendid}.

\paragraph{Neural Decompilation}
Early neural decompilation work
focused on generating \textit{readable}
code from short snippets, rather than semantically correct
code~\cite{8330222}.
Error-repair techniques 
can be used to address correctness ~\cite{Katz2019} and
~\cite{Fu2019}.  However, these works
only explore assembly generated from
synthetic and restricted datasets,
and only from small snippets.

Cao~et~al.~\cite{cao2022boosting}
lifts binaries into intermediate representations, but is only
evaluated against synthetic C programs.
An alternative evolutionary approach based on genetic algorithms is explored in
\cite{schulte2018evolving},
but is only evaluated on 19 programs. 

\paragraph{Binary-Source Code Matching} Gui et al. \cite{matching} propose XLIR leverage contextual embeddings of low-level representations (LLVM's intermediate representations) to find potential matches between pairs of code snippets in potentially different programming languages. While not generative (i.e., they don't generate new code snippets), XLIR is effective at tasks such as clone connection

\paragraph{Neural Code Translation}
Since the advent of sequence-to-sequence models \cite{10.5555/2969033.2969173}, neural machine translation 
has been applied to programming language translation \cite{tree2tree,drissi2018program} often using data from coding websites~\cite{Lu2021a}, and
also in unsupervised settings \cite{lachaux2020unsupervised,Roziere2020,
DBLP:journals/corr/abs-1902-01313, DBLP:journals/corr/abs-2006-03511}.
Other tasks range from 
 code style detection \cite{Pizzolotto2021}, generating accurate variable names \cite{lacomis2019dire}, correcting syntax errors and bugs \cite{santos2018syntax, campbell2014syntax,
hong2021fix},
code completion \cite{DBLP:journals/corr/abs-1905-08325} and program synthesis~\cite{Austin2021} to API recommendation~\cite{Kang2021}, specification synthesis \cite{mandal2023large}
and full-scale code migration~\cite{lachaux2020unsupervised}. 
Regarding low-level code, in \cite{guo2022enabling} they target the generation of LLVM IR from C while
\cite{armengol-estape2021learning} targets the generation of x86.

%% file: conclusion.tex
\section{Conclusion and Future Work}
\label{sec:con}
We present \name{},  a small language  neural decompiler trained on real-world code. \name{} is the first decompiler to combine supervised training and type inference-based program analysis, allowing decompilation of code with external
type and function declarations.
This paper delivers the first neural decompiler portable across ISAs and code optimization levels. We conduct a large-scale evaluation on 4,000 functions against a state-of the art, industrial strength
decompiler, Ghidra, and a general large language model, ChatGPT. Across two benchmark suites, two ISAs and two different types of optimized binaries,
\name{} provides superior performance in terms of both correct decompilation and readability.

Future work should increase the scope of decompilation to larger program units, potentially exploring the use of pre-training
and program repair to improve accuracy. Longer-term, it would be interesting to investigate how learnable and analytic approaches could be  best  integrated to deliver both portable and correct decompilation. More generally, we believe the combination of neural approaches and analytic approaches such as type inference should be further explored.

%% file: appendix.tex
\appendix


The accompanying artifact \cite{slade_authors_2023_10205121} contains the instructions, license, model weights, tokenizers, evaluation data, and code required to evaluate \name{} and the baselines. However, there are certain elements we cannot provide (e.g., BTC's weights, and ChatGPT API keys). 

Additionally, we warn the user that since the IO evaluation requires the host to execute potentially unsafe code, it is advised to execute it in a sandbox at their own risk. If the host doesn't match the expected dependencies and configuration we used, the IO evaluation might fail or yield incorrect results. 

\subsection{Running \name{}}

Executing \name{}'s neural component using the accompanying artifact \cite{slade_authors_2023_10205121} can be as easy as running:
\begin{verbatim}tokenizer.decode(model.generate(
  tokenizer.encode(ASSEMBLY)))
\end{verbatim}
using Huggingface's Transformers Python API.

Nevertheless, the type inference engine and the IO evaluation require additional complexity and dependencies. Assuming a Unix (ideally Ubuntu) machine:

\begin{enumerate}
    \item Preparing the Python virtual environment and dependencies: 
    \begin{verbatim}
python3.8 -m venv venv
source venv/bin/activate
pip install torch==1.10.2
  transformers==4.22.2
pip install -r requirements.txt
\end{verbatim}
\item Installing the PsycheC type inference engine and other system dependencies for the evaluation (assuming Ubuntu):
\begin{verbatim}

sudo apt-get install haskell-stack
   python-is-python3
   gcc-aarch64-linux-gnu 

git clone
 https://github.com/jordiae/psychec
cd psyched;
git checkout original; mkdir build
cd build
cmake ..; make
cd ..; cp build/psychecgen .
cd ..
\end{verbatim}
\item Running \name{} IO evaluation: 
\begin{verbatim}
python new_evaluate_art.py 
\end{verbatim}
The script is parameterized with the following global variables:
\begin{itemize}

\item
  BATCH\_SIZE.
\item
  INFER\_DEPS: Whether to use type inference.
\item
  EVAL\_DATA\_PATH: which benchmark to evaluate on. ExeBench:
  \texttt{eval\_data/exebench}. Synthesis benchmark:
  \texttt{eval\_data/full\_short\_with\_vrecip.json}
\item
  DIRECTION: Which direction to evaluate (the right model will be
  automatically set accordingly).
\end{itemize}

\end{enumerate}

\subsection{Running the baselines}

Similarly, we provide scripts to run the baselines:

\begin{enumerate}
    \item Removing the type inference: Run the same \texttt{new\_evaluate\_art.py} script with the Python global variable \texttt{INFER\_DEPS} set to \texttt{False}.
    \item ChatGPT: Run
    \begin{verbatim}
OPENAI_API_KEY="MY_KEY"
  python new_evaluate_art_oai.py
\end{verbatim}
Unfortunately, we cannot guarantee that the user will get the same outputs as we originally obtained, as ChatGPT is periodically updated.
\item Ghidra: We use the headless version in \url{https://gitlab.com/CinCan/tools/-/tree/master/stable/ghidra-decompiler}, which requires Docker:
\begin{verbatim}
docker image pull
  cincan/ghidra-decompiler
\end{verbatim}

Then run \texttt{new\_evaluate\_art\_ghidra.py}.

\item BTC is expected to be in the \texttt{btc/} directory. However, we cannot
directly provide access to its weights (we got them on request to the
authors but can't redistribute; the user willing to reproduce these
results should request the checkpoints to the corresponding authors).

As an additional dependency for BTC:

\begin{verbatim}
pip install fairseq==0.12.2
\end{verbatim}

Since it's not designed to predict the function signature types, we help it by
prepending the ground truth function signature to its output. Note that BTC only works with `s\_c-c' (i.e., x86 O0).
\end{enumerate}

While the evaluation scripts emit some summary statistics, to obtain the final results one needs to execute the \texttt{plot\_evals.py} script, which generates the plots.

Additionally, we provide output samples for the different models in the \texttt{output/} directory.

Please see the \texttt{README} file attached to the artifact for a complete reference on the scripts and data included in the artifact.